\documentclass[pra,preprint,aps,superscriptaddress,showpacs]{revtex4}
\usepackage{amsmath,amssymb,latexsym,amsfonts}
\usepackage{epsfig}
\begin{document}
\title{Simulation of guiding of multiply charged projectiles through
insulating capillaries}
\date{\today}

\author{K. Schiessl}
\affiliation{Institute for Theoretical Physics, Vienna University of 
Technology, Wiedner Hauptstra\ss e 8-10, A--1040 Vienna, Austria}
\author{W. Palfinger}
\affiliation{Institute for Theoretical Physics, Vienna University of 
Technology, Wiedner Hauptstra\ss e 8-10, A--1040 Vienna, Austria}
\author{K.\ T\H{o}k\'{e}si}
\affiliation{Institute of Nuclear Research of the Hungarian Academy of 
Sciences, (ATOMKI), H--4001 Debrecen, P.O.Box 51, Hungary}
\affiliation{Institute for Theoretical Physics, Vienna University of 
Technology, Wiedner Hauptstra\ss e 8-10, A--1040 Vienna, Austria}
\author{H. Nowotny}
\affiliation{Institute for Theoretical Physics, Vienna University of 
Technology, Wiedner Hauptstra\ss e 8-10, A--1040 Vienna, Austria}
\author{C. Lemell}
\email[Author, correspondence should be sent to: ]{lemell@concord.itp.tuwien.ac.at}
\affiliation{Institute for Theoretical Physics, Vienna University of 
Technology, Wiedner Hauptstra\ss e 8-10, A--1040 Vienna, Austria}
\author{J. Burgd\"orfer}
\affiliation{Institute for Theoretical Physics, Vienna University of 
Technology, Wiedner Hauptstra\ss e 8-10, A--1040 Vienna, Austria}

\begin{abstract}
Recent experiments have demonstrated that highly charged ions can be guided through insulating nanocapillaries along the direction of the capillary axis for a surprisingly wide range of injection angles. Even more surprisingly, the transmitted particles remain predominantly in their initial charge state, thus opening the pathway to the construction of novel ion-optical elements without electric feedthroughs. We present a theoretical treatment of this self-organized guiding process. We develop a classical trajectory transport theory that relates the microscopic charge-up with macroscopic material properties. Transmission coefficients, angular spread of transmitted particles, and discharge characteristics of the target are investigated. Partial agreement with experiment is found.
\end{abstract}

\maketitle

\section{Introduction}
The transmission of multiply and highly charged ions (HCI)
through nanocapillaries has recently been employed as a tool to study the interaction with surfaces, specifically, the internal walls of the capillary. The original motivation was to delineate the initial state of hollow-atom formation at large distances from the surface \cite{nino,burg,tok} information hardly accessible by scattering at flat surfaces. Image attraction and close collisions tend to erase the memory on this early state of ion-surface interaction when the ion suffers close encounter as it is either reflected from the topmost layer of the surface or penetrates into the bulk. A previous first step to access information on the early stages of the neutralization process was the usage of ``stepped'', or terrace-decorated, surfaces \cite{kimura88}. Capillaries provided an attractive alternative as the ion can escape the capillary prior to a close encounter with the surface, the hollow atom can be directly spectroscopically investigated \cite{nino,mori}. Initial investigations focused on metallic capillaries with typical radii of $r$ = 50 to 100 nm (``nanocapillaries'') and a length $L$ of about one $\mu$m, thus featuring an aspect ratio of the order of 1:20 and geometric opening angles $\theta_0 < 3^\circ$. Incidence angles $\theta_{in}$ larger than $\theta_0$ therefore necessarily lead to the impact of all projectiles on the surface and, consequently, to the destruction of the hollow atom (or hollow ion).

More recently, capillaries through insulating foils (PET or ``Mylar'', \cite{stol02}) and SiO$_2$ \cite{vik} with aspect ratios $\sim 1:100$ have been studied in several laboratories \cite{schuch04,aum04,kanai}. Unexpectedly, considerable transmission probabilities for projectiles in their initial charge state were measured for incidence angles as large as $\sim 20^\circ$. Apparently, ions are guided along the capillary axis with a spread (FWHM) of $\Delta \theta_{out}$ of several degrees for mylar \cite{stol02} but close to geometric opening $\theta_0$ for SiO$_2$ \cite{schuch04}. Keeping the initial charge state, contrary to the expected neutralization upon approach of the internal capillary surface, suggests that the ions bounce off the walls at distances larger than the critical distance $R_c \approx \sqrt{2Q}/W$ \cite{burg}: ($Q$: charge state; $W$: workfunction of capillary). Key to this process is the charging up of the internal insulator walls due to preceding ion impacts. Ion guiding through the capillary ensues as soon as a dynamical equilibrium of self-organized charge up by the ion beam, charge relaxation, and reflection is established. 

A theoretical description and simulation of this process poses a considerable challenge in view of the widely disparate time scales simultaneously present in this problem:

\begin{enumerate}
\item The microscopic charge-up and charge hopping due to the impact of individual ion impact takes place on a time scale of sub-$fs$ to $fs$ with typical hopping time $\tau_h < 10^{-15}$ s. 
\item The transmission time $\tau_t$ of a projectile ion through the capillary for typical ion energies of $\approx 200$ eV/u is of the order of $\tau_t \approx 10^{-10}$ s.
\item Typical average time intervals $\overline{\Delta t}$ between two subsequent transmission (or impact) events in the same capillary are, for present experimental current densities of nA/mm$^2$ of the order of $\overline{\Delta t} \approx 0.1$ s.
\item Characteristic (bulk) discharge times $\tau_b$ for these highly insulating materials, can be estimated from conductivity data \cite{mylar} to typically exceed $\tau_b \gtrsim 10^3$ s and can even reach days.
\end{enumerate}

As this multi-scale problem spans a remarkable 18 orders of magnitude, a fully microscopic ab initio simulation covering all relevant scales is undoubtedly out of reach. The more modest goal of the present approach is therefore to develop a simulation that allows to interrelate the microscopic description of ion-surface impact with macroscopic properties of charge-up and transport. Aim is therefore to employ in our simulation only parameters that are not freely adjustable but can, at least as far as their order magnitude is concerned, be deduced from data for macroscopic material properties of the nanocapillary material. Specifically, the bulk discharge time $\tau_b$ as well surface charge diffusion constant $D_s$ will be estimated from surface and bulk conductivity data for mylar \cite{mylar}. 

The present approach represent a mean-field classical transport theory \cite{deiss05} based on a microscopic classical-trajectory Monte Carlo (CTMC) simulation for the ion transported, self-consistently coupled to the charge-up of and charge diffusion near the internal capillary walls. As a limiting case we also consider a simplified rate-equation approach. We find partial agreement with experimental data. We conclude this paper by pointing to future experimental and theoretical challenges to be overcome. 

\section{Method of simulation}

\subsection{The Scenario}
\begin{figure}
\epsfig{file=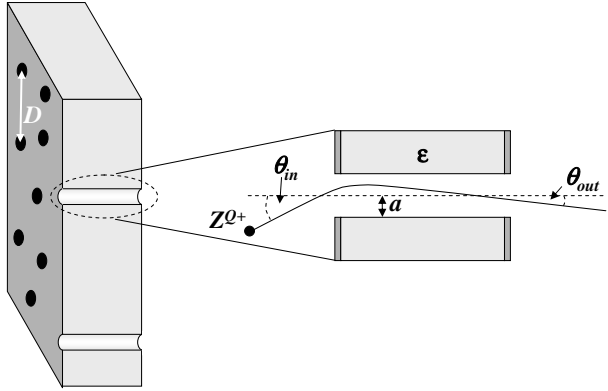,width=8cm,clip=}
\caption{Ion-nanocapillary interaction, schematically: array of nanocapillaries oriented along the surface normal with close-up of an individual capillary. An insulating material (PET)
with dielectric constant $\varepsilon$ is covered on both sides with gold layers (dark shaded) preventing charge up of the target during experiment. Capillaries with radius $a=50$ nm and $L=10\,\mu$m are typically $D=500$ nm apart. Projectiles enter and exit the capillary under angles $\theta_{in}$ and $\theta_{out}$ with respect to the capillary axis. The capillary axis is either normal to the surface or Gaussian distributed with $\Delta\theta_a\lesssim 2^\circ$ (FWHM).}
\label{schema}
\end{figure}

The simulations presented below incorporate the features of the following underlying scenario: an ensemble of ions of charge state $Q$ are incident relative to the surface normal of the nanocapillary foil with angle $\theta_{in}$. The axis of the capillary with radius $a$ is assumed to be either aligned with the surface normal (Fig.\ \ref{schema}) or Gaussian distributed around the orientation of the normal with width $\Delta\theta_a$.
The lateral inter-capillary spacing $D$ is of the order of $D\approx10 \, a$ such that, on an atomic level, inter-capillary interaction effects can be neglected. Mesoscopic field effects resulting from the charge-up of the ensemble of capillaries will, however, be taken into account. The entrance and exit surfaces are covered by a $30$ nm metallic gold layer in order to avoid macroscopic charge-up of the capillary foil. The insulating material on the inside is characterized by a dielectric constant $\varepsilon=3.3$. In view of the low projectile velocities $v_p$ for which guiding is studied, $v_p \lesssim 0.1$ a.u., the static limit $\varepsilon (\omega \rightarrow 0) = \varepsilon$ of the dielectric function $\varepsilon (\omega)$ is appropriate to characterize the linear response. About 4~\% of the entrance surface is covered by capillary openings. Typical current densities are $\approx$ nA/mm$^2$ and the beam spot size is of the order of $\approx$ mm$^2$ simultaneously illuminating $\approx 10^6$ adjacent nanocapillaries. Ions entering the nanocapillary will be attracted by the image field
\begin{equation}
\label{eq:1}
\vec{F}_{im} = \frac{Q}{4 |d^2 |} \frac{\varepsilon - 1}{\varepsilon + 1} \, \hat{r} \, ,
\end{equation}
where $d$ is the distance to the capillary wall. Therefore, even ions entering the capillary with angles $\theta_{in}$ smaller than the geometric aspect ratio $A$,
\begin{equation}
\label{eq:2}
\tan\theta_{0} \approx \theta_0 = A = a/L
\end{equation}
will eventually experience close encounters with the surface. Therefore, the effective angle for transmission would be smaller than given by Eq.\ \ref{eq:2}, unless guiding effects become operational. Ions reaching the surface will undergo a charge transfer sequence yielding a neutralized projectile and deposite a number $Q_{eff}$ of positive charges with $Q_{eff} \gtrsim Q$ at the surface. In our simulation, we use $Q_{eff} = Q$ for the microscopic charge up per ion impact at the capillary wall above all because the secondary electron emission coefficient $\gamma$ for impact of slow ions at insulator surfaces is small $\gamma \lesssim 1$ \cite{aum04}. We have, however, tested the influence of larger $\gamma$ by performing simulations in which up to $2Q$ positive charges and $Q$ electrons with energies up to 50 eV were produced. Trajectories of the latter were followed and these electrons were allowed to neutralize positive charges on the capillary wall. No significant differences between simulations for different $\gamma$ could be found.

The charged-up micro-patch will undergo surface diffusion with diffusion constant $D_s$ as well diffusion into the bulk with diffusion constant $D_b$. Bulk diffusion is extremely slow for highly insulating materials while the surface diffusion towards the grounded metallic layers will be a factor $\simeq 100$ faster, thus governing the overall discharge process. Self-organized guiding sets in when a dynamical equilibrium between charge-up by a series of ion impacts at internal walls with an average frequency $(\overline{\Delta t})^{-1} \approx 5$ Hz and the charge diffusion is established such that the electrostatic repulsion prevents further impacts. The ion is reflected at distances from the wall larger than the critical distances from the surface for electron capture \cite{burg},
\begin{equation}
\label{eq:3}
R_c = \sqrt{2Q}/ W \, . 
\end{equation}
The wall forms then an effective mesoscopic ``trampoline'' for subsequent ions and guides the projectile towards the exit.

\subsection{Projectile Trajectories}
We consider the propagation of $1 \leq n \leq N$ ions through  $1 \leq m \leq M$ nanocapillaries. The calculation of the $n^{th}$ ionic trajectory through the $m^{th}$ capillary proceeds by solving Hamilton's equation of motion
\begin{eqnarray}
\label{eq:4}
\dot{\vec R}_n^{(m)} &=& \frac{\partial H}{\partial \vec{P}_n^{(m)}} \qquad (n = 1, ...N; m = 1, ...M)\\
\label{eq:5}
\dot{\vec P}_n^{(m)} &=& - \frac{\partial V}{\partial \vec{R}_n}
\end{eqnarray}
with random initial conditions subject to the energy constraint
\begin{equation}
\label{eq:6}
\frac{P_n^2 (t = - \infty)}{2m} = E_n \, .
\end{equation}
The non-trivial aspect of this classical-trajectory Monte Carlo simulation is that memory effects must be built into the potential $V$ (Eq.\ \ref{eq:5}). The force experienced by the $n^{th}$ ion depends on the history of all previous trajectories $\{R_{n'}^{\{m'\}}\}$ through the same $(m = m')$ capillary as well as on the ensemble of neighboring charged-up capillaries $m \neq m'$ via the mean field $F_{mean}$. It furthermore depends on the surface and bulk diffusion constants $D_{s,b}$ of the material. By deducing values for $D_{s,b}$ from macroscopic conductivity data and considering them to be independent from the stage of charge-up, we treat the discharge process in linear response, i.e.\ we neglect non-linear processes discussed recently \cite{stol03,stol,schiessl}. Details of the determination of $V$ will be discussed below. 

\begin{figure}
\epsfig{file=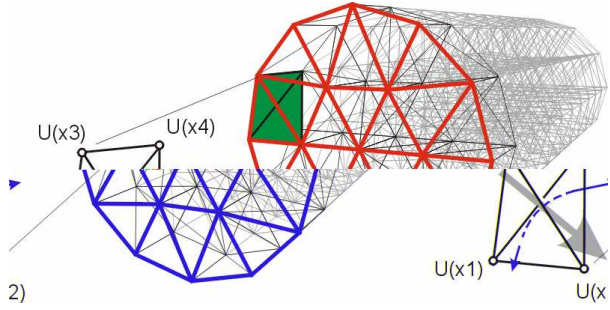,width=8cm,clip=}
\caption{Algorithm for propagation through the capillary, schematically. The capillary is divided into tetrahedrons through which the trajectory is propagated analytically (cf.\ text).}
\label{fig2}
\end{figure}

Solving Eqs.\ \ref{eq:4} and \ref{eq:5} is, while straight-forward in principle, quite demanding in view of the disparate time scales and ensemble sizes involved. We have therefore developed an algorithm that approaches the accuracy of conventional Runge-Kutta methods while considerably increasing the computing efficiency. First, the capillary is divided into three-sided prisms which are further subdivided into three tetrahedrons (see Fig.\ \ref{fig2}).
The force (Eq.\ \ref{eq:5}) is calculated before the simulation of a given trajectory only at the corners of the tetrahedrons. The starting point of the trajectory is determined from a uniform probability distribution over the entrance surface of the capillary, the entrance angle $\theta_{in}$ is selected from a Gaussian distribution centered around the nominal entrance angle. Its full width at half maximum (FWHM $\Delta\theta_{in}\approx 0.5^\circ$) was taken from experimental parameters \cite{stol02}. The propagation is calculated in three steps: first, the point of entrance into a tetrahedron is determined. Next, the average electric field calculated from the potential (Eq.\ \ref{eq:5}) in this structure is determined and, as the last step, the exit point is derived from the \textit{analytical} propagation of the trajectory within the tetrahedron. This exit point becomes the starting point of the trajectory within the next tetrahedron and so forth (see Fig.\ \ref{fig2}). From comparison between our method and a fourth-order Runge-Kutta simulation we have determined the optimal size of the tetrahedrons. We find a significant speed-up (up to a factor 100) while preserving accuracy to a good degree of approximation. 

The propagation of the trajectory is stopped if one of the following two requirements is met: Either the projectile reaches the exit surface of the capillary at which point the exit angle for this trajectory is calculated and the transmission probability of the ensemble is updated or the trajectory is not sufficiently deflected by the field of surface charges and hits the surface. The point of impact is determined and serves as starting point of the simultaneously performed  calculation of the charge diffusion. Diffusion of the deposited $Q$ charges is determined for a time interval $\Delta t$ between two the subsequent trajectories $(n,~n + 1)$ enter the same capillary $m$. We have performed simulations both for Poissonian distributed intervals with mean $\overline{\Delta t}$ as well as for fixed intervals  $\Delta t = \overline{\Delta t}$. The results are indistinguishable which is not surprising as $\overline{\Delta t}$ is 9 orders of magnitude larger than $\tau_t$.

\subsection{Calculations of the Fields Inside the Capillaries}
The forces governing Eq.\ \ref{eq:5} are determined from the electric fields $\vec{F}$ resulting from the charge-up as well as the image field. 
\begin{equation}
\label{eq:7}
\dot P_n^m=Q \vec{F} = Q\left(\vec F_{im}+\vec F_{wall}+\vec F_{mean}\right)
\end{equation}
with $\vec{F}_{im}$ given by Eq.\ \ref{eq:1}, $\vec{F}_{coul}$ represents the electric field due to the charge-up of the wall of the $m^{th}$ capillary, and $\vec{F}_{mean}$ is the mesoscopic field resulting from the charge-up of the entire ensemble of capillaries.

Calculation of the charge distribution makes use of the macroscopic properties of these highly insulating materials \cite{mylar}. Starting point is the bulk conductivity of Mylar
$\sigma_b\cong 10^{-16}\, \Omega^{-1}$m$^{-1}$. The surface conductivity $\sigma_s$ is larger by a factor of about 100 \cite{mylar}. From the Einstein relation
\begin{equation}
\label{eq:8}
\sigma_{b,s}=\frac{ne^2}{kT}D_{b,s}
\end{equation}
and using the experimentally determined value for the amount of charge carriers in Mylar $n \cong 10^{18}$ m$^{-3}$ \cite{lilly68}, $D_b$ should be of the order of $10^{-17}$ m$^2$s$^{-1}$ and $D_s \approx 10^{-15}$ m$^2$s$^{-1}$. We use $D_b=2 \cdot 10^{-15}$ m$^2$s$^{-1}$ and $D_s=100 \cdot D_b$ unless otherwise stated.
The disparity of surface and bulk diffusion will be key for the understanding of the charge-up and discharge characteristics. The point to be noted is that for such low conductivities and large uncertainties in the effective number of carriers present, Eq.\ \ref{eq:8} can only serve for an order of magnitude estimate. We will later check on the consistency of these estimates by comparison with results of our simulation. After deposition of $Q$ charges per impact, charges are assumed to undergo an unbiased random walk resulting in the diffusive spread of the charges. For simplicity and to avoid the introduction of any free parameter, modification of the random walk by mutual Coulomb repulsion is neglected. Accordingly, the probability distribution for the charge distribution in the surface a time interval $\Delta t$ after impact is given by \cite{statmech}
\begin{equation}
\label{eq:9}
P_s (\rho, \Delta t) = \frac{Q}{4\pi D_s\Delta t}\exp\left(\frac{-\rho^2}{4D_s\Delta t}\right)\; ,
\end{equation}
where $\rho$ is the distance on the surface. Likewise, the distribution for bulk diffusion in 3D is given by
\begin{equation}
\label{eq:10}
P_b (u, \Delta t)= \frac{Q}{(4\pi D_s\Delta t)^{3/2}}\exp\left(\frac{-u^2}{4D_b\Delta t}\right)
\end{equation}
with $u$ the distance in 3D. Since surface diffusion is much faster and effectively controls the discharging at the boundary to the metallic layer, we treat surface diffusion explicitly using Eq.\ \ref{eq:9}. By contrast, we simulate bulk diffusion after the random walk into the bulk by a switch-off of the charge with a time constant $\tau_b$ estimated from $D_b$ as 
\begin{equation}
\label{eq:11}
\tau_b =\frac{l_b^2}{D_b}
\end{equation}
where we use as characteristic hopping length $l_b$ into the bulk half of the mean inter-capillary distance (i.e.\ $l_b=D/2 \approx 5a$ \cite{stol02}).

The wall potential seen by the projectile entering the capillary is expressed by
\begin{equation}
\label{eq:12}
V_{wall}(\vec r,t)=\int\limits_{surface} da\; \frac{\sigma(\vec r',t)}{|\vec r-\vec r'|}
+\sideset{}{'}\sum_{\{j\}}\frac{\exp(-(t-t_j)/\tau_b)}{|\vec r-\vec r_j|}.
\end{equation}
In the first term of Eq.\ \ref{eq:12}, the surface charge density is given in terms of the sum over all surface diffusion distributions resulting from impacts at times $t_k<t$,
\begin{equation}
\label{eq:13}
\sigma (\vec r, t) = (1-P_b)\sum_{\substack{k\\(t_k<t)}} P_s(\vec r-\vec r_k,t-t_k).
\end{equation}
The second term in Eq.\ \ref{eq:12} describes the exponentially decaying Coulomb interaction of the fraction of those $\{j\}$ charges disappearing into the bulk at time $t_j$ and at position $\vec{r}_j$ with probability $P_b$. From Eq.\ \ref{eq:12}, the field $\vec{F}_{wall} = - \Delta V_{wall}$ can be determined. 

Finally, the mean field $\vec{F}_{mean}$ takes into account collective field effects of an entire ensemble of nanocapillaries approximately $ \approx 10^6$ of which are located within the beam spot. As they get simultaneously charged-up, they generate a mesoscopic electric field that is aligned along the capillary axis. Due to the large distance of the charge patches near the entry side from the exit surface ($\approx 10\, \mu$ m) the ensemble can be viewed as a charged condenser consisting of two metallic plates (the gold layers) with a dielectric material characterized by the dielectric constant $\varepsilon$ in between (Fig.\ \ref{schema}). The mean field can be estimated from the total charge deposited on the capillary walls, the dielectric constant of the insulating material, and the irradiated target area.

\begin{figure}
\epsfig{file=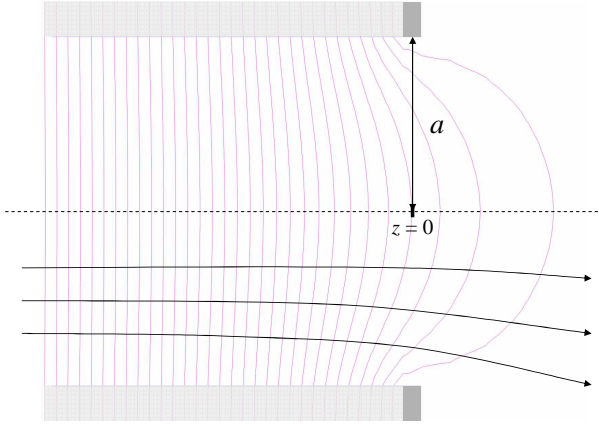,width=8cm,clip=}
\caption{A metal plate with a hole with radius $a$ separates two half spaces. On the
right hand side $(z>0)$ Dirichlet boundary conditions are assumed for $r,z\to\infty$. The boundary condition for $z\to -\infty$ is given by the requirement that the electric field
converges to a homogeneous field $\vec F_0\| \hat z$.}
\label{fig3}
\end{figure}

We have performed simulations of the potential and field $\vec F_{mean}$ near the exit surface using the program POISSON/SUPERFISH \cite{fish}. Equipotential lines from these calculations are shown in Fig.\ \ref{fig3}.
For efficient implementation into the classical transport simulation we also developed a simplified analytic model for $\vec{F}_{mean}$. To this end, the exit plane of the capillary is approximated by a infinitely thin metal plate with a hole of radius $a$. For $z \to -\infty$, a homogeneous field $\vec F_0$ proportional to the amount of charge deposited at the entrance surface along the capillary axis is imposed as boundary condition, near the exit surface $(z = 0)$ Dirichlet boundary conditions are applied. Then, the potential in the half sphere with $z > 0$ is given by \cite{helmut}
\begin{equation}
\label{eq:14}
\Phi^+(r,z)=-\frac{F_0}{\pi}\,[b-z\cdot\arctan (b/z)]
\end{equation}
with
\begin{eqnarray}
\label{eq:15}
b&=&\sqrt{a^2-l^2}\\
\label{eq:16}
l&=&\frac{1}{2}\left[\sqrt{(r+a)^2+z^2}-\sqrt{(r-a)^2+z^2}\,\right].
\end{eqnarray}
The potential in the negative half space is determined by
\begin{equation}
\label{eq:17}
\Phi^-(r,-z)=\Phi^+(r,z)-F_0z
\end{equation}
with the matching conditions at $z=0$
\begin{equation}
\label{eq:18}
-\left.\frac{\partial\Phi^-}{\partial z}\right|_{z\to 0_-}=
-\left.\frac{\partial\Phi^+}{\partial z}\right|_{z\to 0_+}\qquad (r<a).
\end{equation}
The analytic model gives fields $\vec{F}_{mean} = - \vec\nabla\Phi$ in close agreement with the numerically calculated field distribution. We therefore employ Eqs.\ \ref{eq:14} -- \ref{eq:18} in our CTT. The importance of the inhomogeneity of $\vec{F}_{mean}$ lies in its influence on the angular distribution of the guided ions by defocusing the transmitted beam near the exit surface.

\section{Results}
\subsection{Transmission function for ion guiding}
\begin{figure}
\epsfig{file=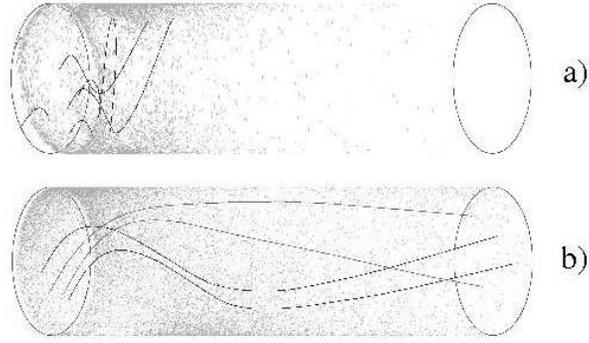,width=8cm,clip=}
\caption{Scatter plot of deposited charges in the interior of an individual capillary and typical trajectories for $\theta_{in}=3^\circ$. a) zig-zag distribution leading to blocking (for $D_s=D_b$); b) patch distribution leading to transmission (for $D_s=100 D_b$). Note that in b) two groups of typical trajectories can be observed, leading to two maxima in the angular distribution (Fig.\ \ref{fig6}). Ions enter from the left, aspect ratio $a/L=100$, not to scale.}
\label{fig4}
\end{figure}

Following the pioneering work of Stolterfoht et al.\ \cite{stol02} several groups studied the ion guiding through Mylar (PET) \cite{schuch04,aum04,kanai} and SiO$_2$ nanocapillaries \cite{vik}. Key finding is the build-up of a self-organized charge distribution that enables the collision-free transmission. Patterns of the self-organized charge distribution are shown in Fig.\ \ref{fig4}.
While Fig.\ \ref{fig4}a shows a typical charge-up condition leading to blocking (electrostatic ``bottleneck''), Fig.\ \ref{fig4}b displays a distribution giving rise to guiding. The bottleneck condition was realized by setting $D_s$ to be equal to the bulk values of $D_b$, corresponding to a slow overall discharge time set here to be $\tau_b \simeq 35$ min. In fact, such a slow discharge time has been suggested by experiment \cite{stol02} for the time interval over which recharging after beam switch-off could be realized. Only in the simultaneous presence of both fast (via surface transport, $D_s$) and slow decay (via bulk transport $D_b$) can guiding be established and maintained. When taking into account both surface and bulk diffusion, assumptions about a non-linear discharging characteristics \cite{stol03,stol,schiessl} need not to be invoked.

\begin{figure}
\epsfig{file=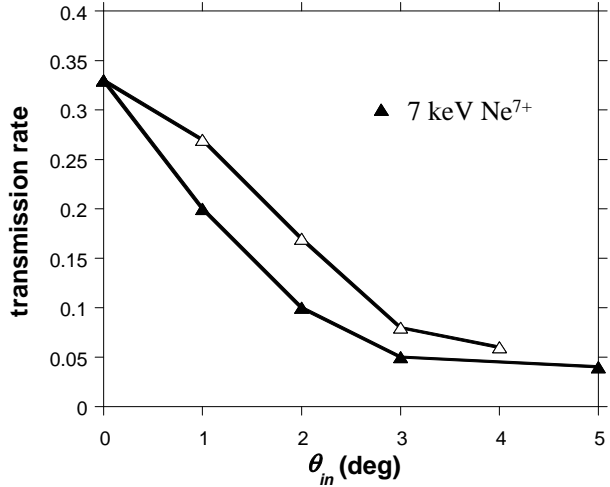,width=8cm,clip=}
\caption{Transmission function (transmission probability) as a function of angle of incidence $\theta_{in}$ relative to the mean capillary axis. Full symbols: present CTT, open symbols: experimental data: 7 keV Ne$^{7+}$ (Vikor et al.\ \cite{schuch04}). Experimental transmission rates have been normalized to CTT results at $\theta_{in}=0^\circ$.}
\label{fig5}
\end{figure}

The resulting transmission probability as a function of the angle of incidence relative to the capillary axis (Fig.\ \ref{fig5}) displays significant transmission for angles well outside the geometric opening angle $\theta_0$. The quantitative agreement with the experimental transmission function is reasonably good considering the discrepancies between different data sets. The latter is, in part, due to the extraordinarily long bulk discharge times (from $D_b$ estimated to be in the range of hours) which makes measurements under reproducible conditions of complete initial discharge a challenge. 

\subsection{Angular distributions}
\begin{figure}
\epsfig{file=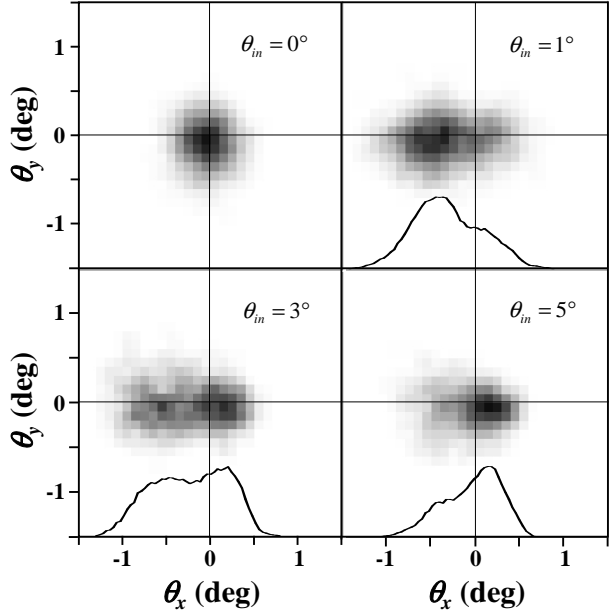,width=8cm,clip=}
\caption{Two-dimensional angular distributions of 3 keV Ne$^{7+}$ ions transmitted through PET capillaries for $\theta_{in}=0^\circ$ ,1$^\circ$, 3$^\circ$, and 5$^\circ$. The $\theta_x$ and
$\theta_y$ directions are defined parallel and perpendicular to the plane of incidence given by the ion beam and the capillary axis, respectively. Solid lines are projections of the distributions on the $\theta_x$-axis.}
\label{fig6}
\end{figure}

The two-dimensional angular distribution of the guided ions displays anisotropic structures (Fig.\ \ref{fig6}) in the $\theta_x$--$\theta _y$ plane where $\theta_x$ is the angle relative to the capillary axis in the ``scattering plane'', the plane spanned by the incident velocity vector and the capillary axis while $\theta_y$ is the angle perpendicular to the scattering plane. The $\theta_y$ distribution normal to the plane of incidence remains almost constant for all angles. Parallel to the plane of incidence $(\theta_x)$ broadening and displacement of the peak from the center of the distribution is found.
This is in agreement with experiments which show a small deviation of the centroid of the scattering distributions towards larger scattering angles for large incidence angles $\theta_{in}$ \cite{stol02}. At certain incidence angles (e.g.\ 3$^\circ$) structures in the distribution (solid lines in Fig.\ \ref{fig6}) become visible which qualitatively agrees with recent experimental findings \cite{aum04}. From the simulation we can delineate the origin of the second maximum as being due to the formation of a small secondary charge patch close to the exit surface which deflects projectiles passing by at close distance. We point out that double peak structures are stable under variation of $D_s$ and the capillary length $L$ although they appear at different angles $\theta_{in}$. It remains to be investigated to what extent double peaks in the data are non-accidental and can be taken as evidence for patch formation near the exit surface. 

\begin{figure}
\epsfig{file=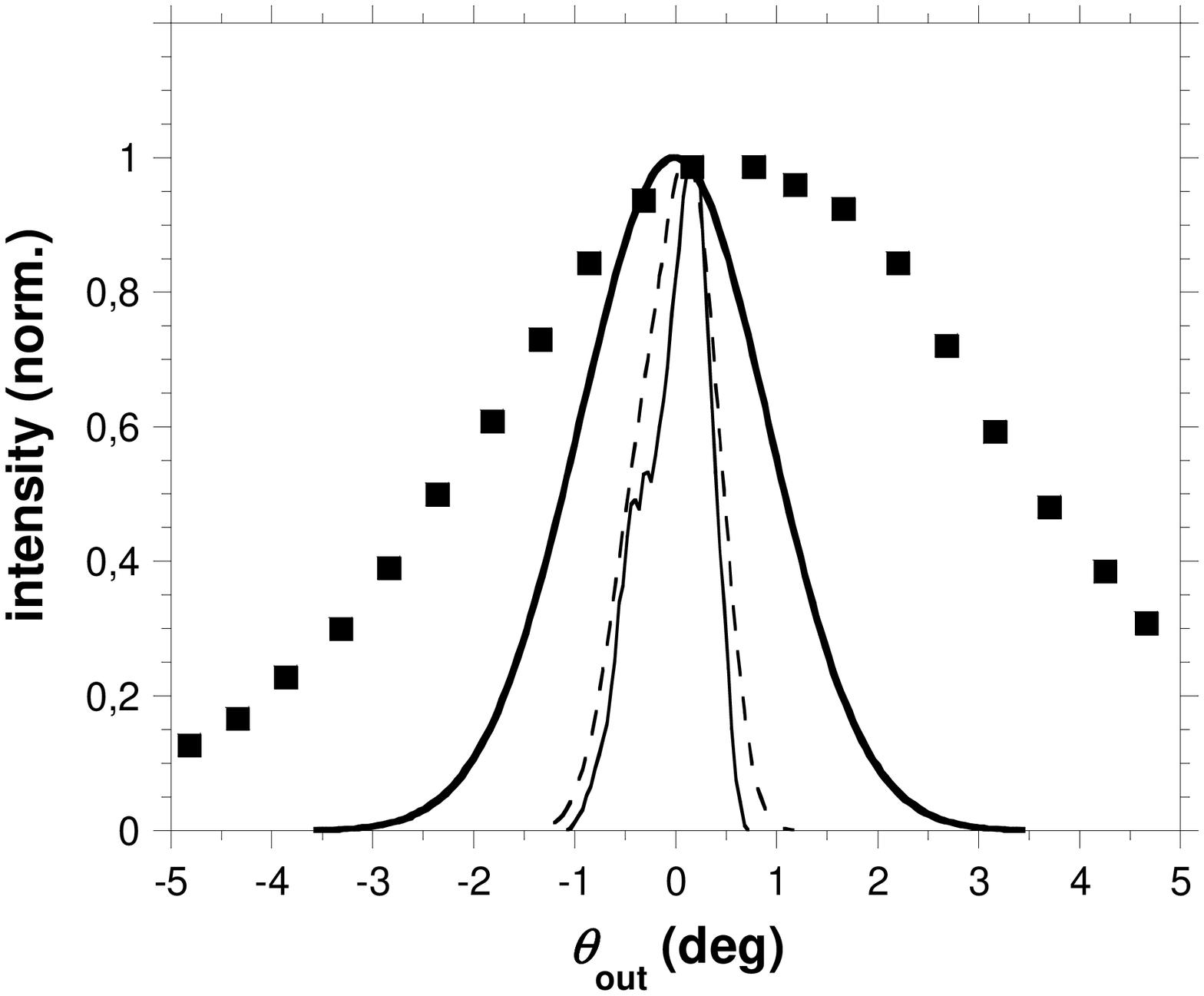,width=8cm,clip=}
\caption{Comparison between experimental (squares \cite{stol02}) and simulated (solid line) angular distributions of exiting Ne$^{7+}$ ions including angular spread of capillary axis and defocussing by $\vec F_{mean}$. Thin solid line: Simulation neglecting defocusing by $\vec F_{mean}$ ($\vec F_{mean}$=0) and spread in capillary axis ($\Delta \theta_a=0$). Dashed line: Simulation including $\vec F_{mean}$ but without spread in capillary axis ($\Delta \theta_a=0$).}
\label{fig7}
\end{figure}

The mean angular spread determined by projecting the data of Fig.\ \ref{fig6} onto the $\theta_x$ axis can be compared with experimental data (Fig.\ \ref{fig7}).
All distributions in Fig.\ \ref{fig7} have been normalized to the same maximum value. The simulated spread $\Delta \theta_{out}$ (FWHM) with $\vec F_{mean}=0$ is close to the geometric angle $\Delta\theta_0$. Including the defocussing effect due to $F_{mean}$, $\Delta\theta_{out}$ increases by $\lesssim 1^\circ$. Limited by the DC dielectric break-down strength of mylar, the defocussing effect of $\vec F_{mean}$ on the total spread $\Delta \theta_{out}$ should not exceed $\approx 2^\circ$. An additional source of spread is the spread in capillary angles. Measurements by Stolterfoht et al.\ indicate an additional spread of up to $\Delta \theta_a \simeq 2^\circ$ due to imperfection in the target preparation can occur \cite{stol04}. Including these contributions, we find $\Delta \theta_{out} \approx 2.5^\circ$ (solid line in Fig.\ \ref{fig7}), which is still somewhat smaller than the value observed in \cite{stol02}. 

\subsection{Transient charging and discharging effects}
\begin{figure}
\epsfig{file=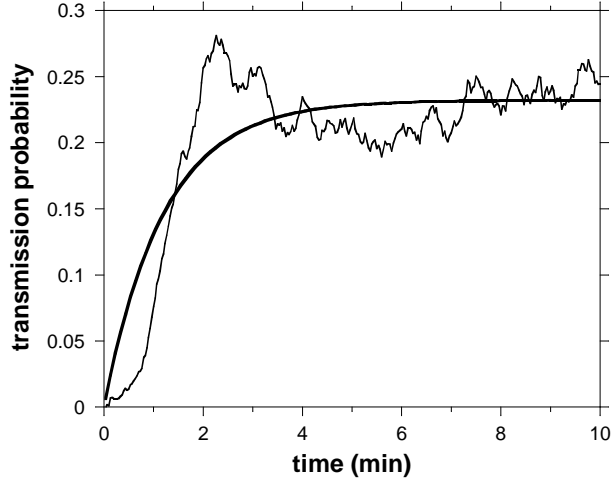,width=8cm,clip=}
\caption{Simulated transmission for 3 keV Ne$^{7+}$ ions with $\theta_{in}=1^\circ$ averaged over 5 (dashed line) and 25 capillaries (full line). The thick line is given by Eq.\ \ref{eq:23} with $\tau_{eff}\approx 1.5$ min.}
\label{fig8}
\end{figure}

\begin{figure}
\epsfig{file=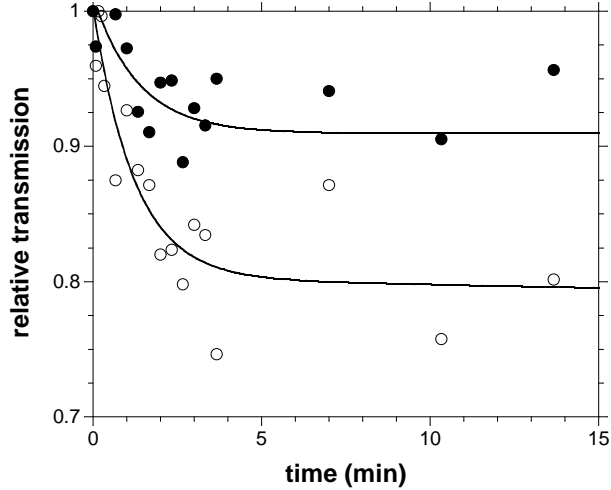,width=8cm,clip=}
\caption{Simulated transmission for 3 keV Ne$^{7+}$ ions with $\theta_{in}=1^\circ$ (filled circles) and $\theta_{in}=5^\circ$ (open circles) after the ion beam has been turned off at time $t_0=0$. Data has been normalized to the transmission at $t=t_0$. Lines given by Eq.\ \ref{eq:26}.}
\label{fig9}
\end{figure}

Establishing the dynamical equilibrium necessary for guiding takes a finite time interval determined by the incoming ion current density and diffusion speed. Studying the transient behavior as the ion beam is switched on and switched off provides additional information on the patch formation causing guiding. Fig.\ \ref{fig8} displays the time-dependent transmission 
probability after the beam is switched on at $t = 0$ assuming that the nanocapillary was initially completely discharged. Note that the latter requirement is difficult to assess and to realize in the experiment. This may explain the observation of different transient behaviors. The simulated transient discharge behavior for 3 keV Ne$^{7+}$ with $\theta_{in}=1^\circ$ and $\theta_{in}=5^\circ$ is shown in Fig.\ \ref{fig9}. It is probed by first switching off the ion
beam at $t=t_0$ and switching it on again after a time interval $\delta t=t-t_0$ for a sufficiently short time interval in order not to disturb the discharge process. Data points are normalized to the transition rate at $t=t_0$ and represent an average transmission over 10 capillaries.

The charging and discharging behavior can be described in terms of a simple analytic rate equation model which is a generalization of the model proposed in \cite{stol03,stol}. To this end we approximate the full surface diffusion including the charge absorption at the nearby gold layer by an effective surface discharge time which we estimate, in analogy to the bulk discharge time (see Eq.\ \ref{eq:11}), by
\begin{equation}
\label{eq:19}
\tau_s = a^2/D_s\, .
\end{equation}
The apparent $\tau_s$ is expected to depend on the angle of incidence $\theta_{in}$. On the one hand, the extension of the charge patch from the gold layer at the entrance scales as $\sim 1/\tan(\theta_{in})\approx 1/\theta_{in}$, on the other hand, projectiles with a larger energy component normal to the surface $E_\perp=E\sin^2\theta_{in}$ require a larger amount of total charge deposited on the capillary surface to be deflected along the capillary axis. The latter requirement is directly reflected in the later onset of transmission for $\theta_{in}=5^\circ$ as compared to the switch-on in the case of $\theta_{in}=1^\circ$ depicted in Fig.\ \ref{fig8}. Taking these effects into consideration we expect for $\tau_s$ a weak dependence on $\theta_{in}^{-x};\, x<1$.

The balance equation for charge $q_s(t)$ deposited at the surface of the capillary reads 
\begin{equation}
\label{eq:20}
\dot{q}_s(t) = j_{in} - j_{tr} - j_s
\end{equation}
where $j_{in}$ is the incident current, $j_{tr}$ the transmitted current, and $j_s$ is the current absorbed in the capillary surface. The latter has two contributions, current into the bulk and current along the surface eventually reaching the gold layer.

We set
\begin{eqnarray}
\label{eq:21}
j_s &=& \left( \frac{1}{\tau_s} + \frac{1}{\tau_b} \right) q_s\\
\label{eq:22}
\dot q_b=-j_b &=&\frac{q_s-q_b}{\tau_b}\; .
\end{eqnarray}

Solving Eq.\ \ref{eq:20} for switch-on with initial condition $q_s (t = 0) = 0$, we find 
\begin{equation}
\label{eq:23}
q_s(t) = \tau_{eff}\;(j_{in}-j_{tr})\left( 1-e^{-t/\tau_{eff}}\right)
\end{equation}
where
\begin{equation}
\label{eq:24}
\tau_{eff}^{-1} =\frac{1}{\tau_s} + \frac{1}{\tau_b}\simeq\tau_s^{-1}
\end{equation}
since $\tau_s \ll \tau_b$. For the switch-off at $t = t_0$ we find 
\begin{equation}
\label{eq:25}
q_s (t) =q_s(t_0)\cdot e^{-t/\tau_{eff}}.
\end{equation}
Solving Eq.\ \ref{eq:22} for any initial condition $q_b(t_0)\leq q_s(t_0)$ and using the solution for switch off of the rapidly varying function $q_s(t)$ leads for the total charge-up of the capillary (or, equivalently, the capillary transmission) to
\begin{equation}
\label{eq:26}
q(t)=q_b(t)+q_s(t)=q_0\cdot\left[C\cdot e^{-t/\tau_b}+(1-C)\cdot e^{-t/\tau_{eff}}\right]
\end{equation}
with $C=C(q_s(t_0),\tau_b,\tau_s)<1$, which is a slowly decaying function (time constant $\tau_b$) after a fast initial decay with time constant $\tau_{eff}\approx\tau_s$ as also found in experiment \cite{stol03,stol}. A comparison with the simulation shows that the switch-on transmission closely follows Eq.\ \ref{eq:23}. It should be noted that the delayed onset of transmission cannot be reproduced by a simple rate equations. This reflects the fact that the rate equation does not account for the threshold behavior for the onset of the trampoline effect. Transmission during switch-off mirrors the total discharge (Eq.\ \ref{eq:26}). The simulation shows that the time constant $\tau_s$ is a decreasing function of the incidence angle $\theta_{in}$ as expected. Fits to the simulation give values of $\tau_s\approx 1.5$ min and $\tau_b \geq 35$ min, their order of magnitude consistent with the parameters derived from the macroscopic conductivity data \cite{mylar}. The point to be emphasized is that no non-linear processes have to be invoked to account for the discharge characteristics.

\section{Conclusions}
We have presented a simulation for ion guiding through insulating nanocapillaries within the framework of a mean-field classical transport theory. We combine a microscopic trajectory simulation with macroscopic material properties for bulk an surface conductivities of highly insulating materials (PET, ``Mylar''). Projectiles hitting the inner wall of the insulating material in the early stage of the irradiation deposit their charge on the capillary surface. These
charges diffuse along the surface and, eventually, into the bulk due to the small but finite
surface and bulk conductivities of the insulator. Projectiles entering the capillary at a
later stage are deflected by the Coulomb field of a self-organized charge patch close to the
entrance of the capillary. Invoking this scenario we are able to reproduce a range of
experimental findings, e.g., ion guiding even for large incidence angles,
the temporal decrease of transmission during beam-off times, and, in part, a relatively large
angular spread of the transmitted beam. We have shown, that these results can be
interpreted on the basis of a linear model including transport of deposited charges along the surface without resorting to freely adjustable parameters. Future investigations should address the dependence on material properties. In particular, measurements for other insulating materials (SiO$_2$) should provide for benchmark data and critical tests for the theoretical framework presented here.

\begin{acknowledgments}
We are grateful to N. Stolterfoht (Berlin) and F. Aumayr (Vienna) for fruitful discussions and for making their results available to us prior to publication.

The work was supported by the Hungarian Scientific Research Found: OTKA Nos.\ T038016, T046454,
the grant ``Bolya'' from the Hungarian Academy of Sciences, the TeT Grant No.\ A-15/04,
the Austrian Fonds zur F\"orderung der wissenschaftlichen Forschung, FWF-SFB016 ``ADLIS'' and by the EU under contract No.\ HPRI-CT-2001-50036.
\end{acknowledgments}

\end{document}